\documentclass[twoside,11pt]{article}

\usepackage{caption}
\usepackage{subcaption}
\usepackage[preprint]{jmlr2e}

\usepackage[disable]{todonotes} % Hide notes without deleting them manually.
\presetkeys{todonotes}{inline, backgroundcolor=pink}{}

\usepackage{amsmath}
\usepackage{lastpage}

\jmlrheading{}{2021}{1-\pageref{LastPage}}{5/21; Revised 6/21}{}{x}{Laila Melkas, Rafael Savvides, Suyog Chandramouli, Jarmo M\"akel\"a, Tuomo Nieminen, Ivan Mammarella and Kai Puolam\"aki}

\ShortHeadings{Interactive Causal Structure Discovery in Earth System Sciences}{Melkas, Savvides, Chandramouli, M\"akel\"a, Nieminen, Mammarella and Puolam\"aki}

\firstpageno{1}

\begin{document}

\title{Interactive Causal Structure Discovery in \\ Earth System Sciences}

\author{\name Laila Melkas \email laila.melkas@helsinki.fi \\
\name Rafael Savvides \email rafael.savvides@helsinki.fi \\
\name Suyog H. Chandramouli \email suyog.hc@helsinki.fi \\
\name Jarmo M\"akel\"a \email jarmo.makela@helsinki.fi \\
\addr Department of Computer Science\\
P.O. Box 68 \\
FI-00014 University of Helsinki, Helsinki, Finland
\AND
Tuomo Nieminen \email tuomo.nieminen@helsinki.fi \\
\name Ivan Mammarella \email ivan.mammarella@helsinki.fi \\
\addr Institute for Atmospheric and Earth System Research/Physics \\
P.O. Box 64\\
FI-00014 University of Helsinki, Helsinki, Finland
\AND
Kai Puolam\"aki \email kai.puolamaki@helsinki.fi \\
\addr Institute for Atmospheric and Earth System Research \\
Department of Computer Science\\
P.O. Box 68 \\
FI-00014 University of Helsinki, Helsinki, Finland
}

\editor{ Thuc Le, Jiuyong Li, Greg Cooper, Sofia Triantafyllou, Elias Bareinboim, Huan Liu, and Negar Kiyavash }

\maketitle

\begin{abstract}%   <- trailing '%' for backward compatibility of .sty file
Causal structure discovery (CSD) models are making inroads into several domains, including Earth system sciences. 
Their widespread adaptation is however hampered by the fact that the resulting models often do not take into account the domain knowledge of the experts and that it is often necessary to modify the resulting models iteratively.
We present a workflow that is required to take this knowledge into account and to apply CSD algorithms in Earth system sciences. 
At the same time, we describe open research questions that still need to be addressed.
We present a way to interactively modify the outputs of the CSD algorithms and argue that the user interaction can be modelled as a greedy finding of the local maximum-a-posteriori solution of the likelihood function, which is composed of the likelihood of the causal model and the prior distribution representing the knowledge of the expert user. We use a real-world data set for examples constructed in collaboration with our co-authors, who are the domain area experts.
We show that finding maximally usable causal models in the Earth system sciences or other similar domains is a difficult task which contains many interesting open research questions. We argue that taking the domain knowledge into account has a substantial effect on the final causal models discovered.
\end{abstract}


\section{Introduction}

In the sciences, the analysis of measurements or observations with many variables is a commonly occurring problem. The objective of such an analysis is typically to find relations between variables. The found relations can then be used for different purposes, such as to uncover physical, chemical, and biological processes that manifest themselves in the measurements, to help in designing new experiments that will fill in the gaps in the current knowledge, or to make computational models that can be used to estimate the values of unobserved latent variables.

For any set of measurements, there is almost always prior knowledge which is used. At simplest, the prior knowledge affects the selection of variables: scientists typically choose to include into their analysis only measurements that they think are relevant for the processes of interest. We will argue in this paper that this prior knowledge can and should be used iteratively in a more fine-grained manner than just for variable selection.

This paper is motivated by recent advances in causal modelling (see Section~\ref{sec:related}). 
Over the years, multiple causal structural discovery (CSD) algorithms have been proposed for finding causal structures from purely observational data. However, each of these algorithms make different assumptions about the underlying data generating process regarding, for example, the functional family of the causal relations or the noise distributions. Different algorithmic choices are used to find the causal model that fits the data best; see \citet{runge2019detecting} for a recent review in Earth system sciences. The causal structure is commonly represented using a directed acyclic graph (DAG) of cause-effect relationships between variables. In this paper, we use the terms {\em causal model} and DAG interchangeably. Causal discovery algorithms all work similarly in this context: they take in a set of measurements and they output a causal model of the observations or a class of such causal models. Because the underlying assumptions differ between causal discovery algorithms, it is typical that different algorithms produce different outputs for the same input data~\citep{druzdzel2009role}. Additionally, even if the modelling assumptions in the causal discovery process are correct, insufficient or biased data may result in skewed results.

In this paper, we focus on the domain of Earth sciences. We claim that having the output(s) of the causal discovery process is just the first step of the process of understanding and using the data. Our objective is to point out steps that need to be taken {\em after finding these initial causal models} and to propose such a workflow. In addition to proposing one possible solution for this problem, we also want to review the open research questions that are relevant to interactive causal modelling in Earth system sciences and other similar domains.

\begin{figure}
    \centering
    \includegraphics[width=\textwidth]{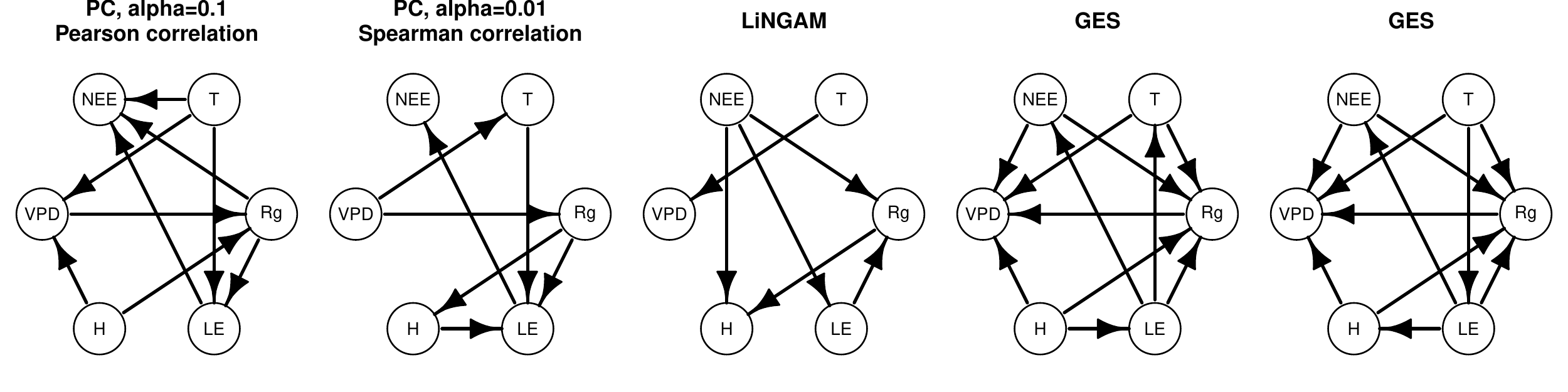}
    \caption{Different algorithms produce different causal graphs for the same data. How can an expert user edit them to incorporate their knowledge? }
    \label{fig:intro_csd}
\end{figure}

To demonstrate the concept, we have run three different causal discovery algorithms on a data set of observations from Hyyti\"al\"a forestry field station in central Finland.  The variables in this data set include daytime measurements of shortwave downward radiation (Rg), air temperature (T), vapour pressure deficit (VPD), sensible heat flux (H), latent heat flux (LE), and net CO$_2$ fluxes, also referred to as net ecosystem exchange (NEE), measured above the tree canopy. The data set and algorithms are described in more detail later in Section~\ref{sec:experiments}.  Outputs of the various algorithms are shown in Figure~\ref{fig:intro_csd}. In theory we would expect to see, for example, the effect of solar radiation (Rg) on air temperature (T) or the effect of solar radiation and temperature on NEE.

However, different algorithms have different outputs for the same input data and it is not clear which of the found models is ``correct'' or how we should proceed from here. We can argue that some of the graphs are more plausible, given what we know of the processes, but it is in practice difficult for an expert to explicitly state their prior knowledge, let alone incorporate it into the causal discovery algorithm. For example, the algorithm may find a relation that temperature causes high solar radiation (an edge pointing from temperature to solar radiation, as in the rightmost graph of Figure~\ref{fig:intro_csd}), while an expert would know that solar radiation causes temperature to rise and not vice versa. While inputting prior knowledge is possible for some of the algorithms and implementations we use, allowing the user to iteratively update their background knowledge into the modelling process or to express uncertainty in the prior information has not been built in. This ambiguity limits the usability of the causal discovery algorithms.

In this paper, we take a Bayesian probabilistic approach to interactive causal structure discovery. We formulate the problem as building a probabilistic model of the data. We assume that the expert's prior knowledge can be characterised by a {\em prior distribution} over all possible causal structures. We show that even if we do not know this prior distribution in advance, we can through interaction with the expert find a causal model that has a local maximum probability in the expert's approximate posterior, that both fits the data and is consistent with what the expert already knows about the phenomena of interest.

\begin{figure}[t]
     \centering
     \begin{subfigure}[b]{0.3\textwidth}
         \centering
         \includegraphics[width=0.7\textwidth]{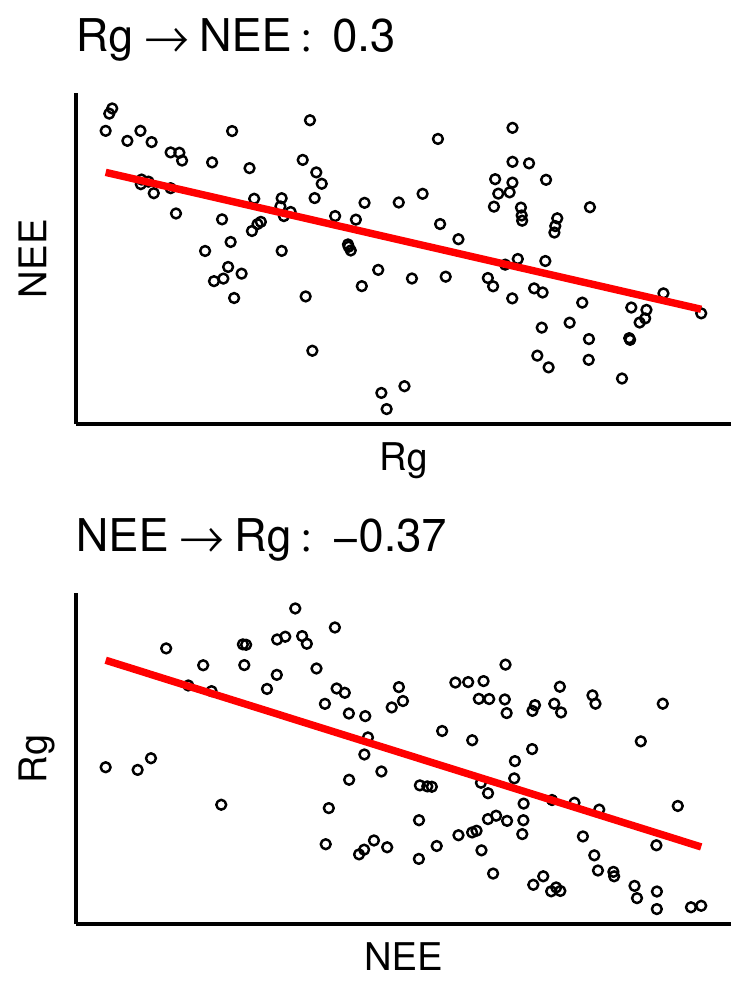}
         \caption{}
         \label{fig:motiv:twovars}
     \end{subfigure}
     \hfill
     \begin{subfigure}[b]{0.3\textwidth}
         \centering
         \includegraphics[width=\textwidth]{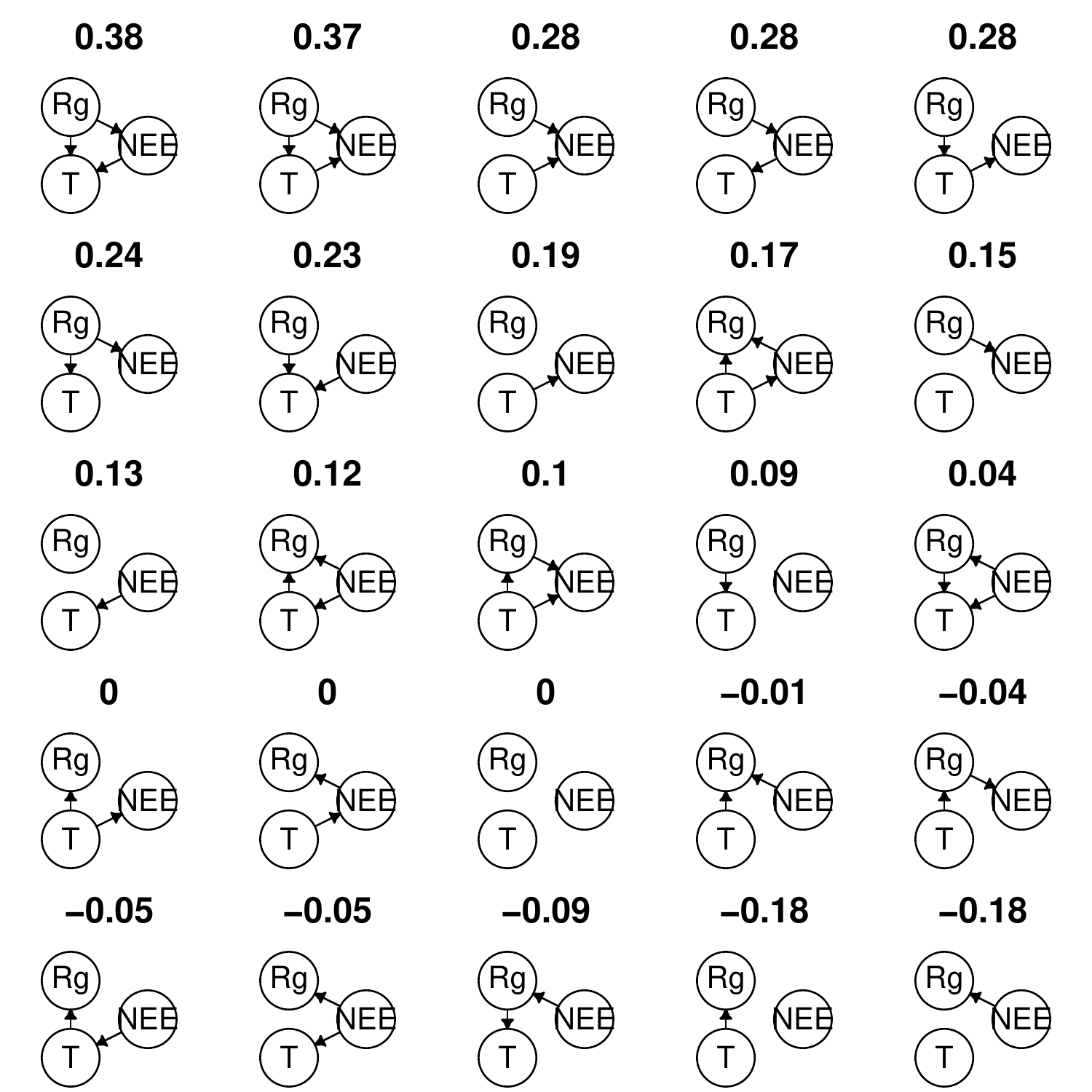}
         \caption{}
         \label{fig:motiv:threevars}
     \end{subfigure}
     \hfill
     \begin{subfigure}[b]{0.3\textwidth}
         \centering
         \includegraphics[width=\textwidth]{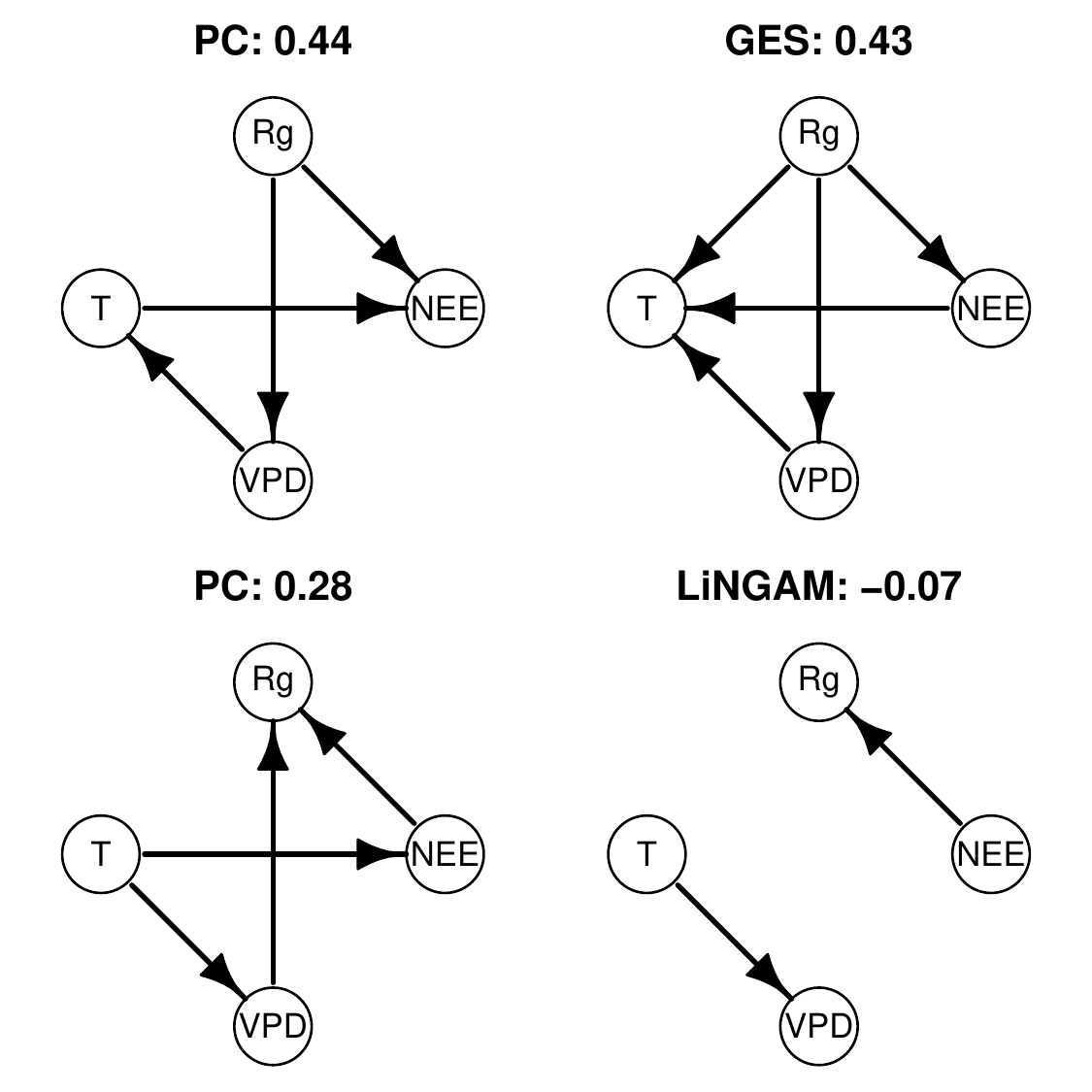}
         \caption{}
         \label{fig:motiv:fourvars}
     \end{subfigure}
    \caption{\label{fig:motiv}
    Motivating example. The space of DAGs increases rapidly with the number of considered variables. An expert cannot manually specify their prior for all possible models. The model score is indicated for each model. See the text for details.}
\end{figure}

As a motivating example, consider finding the causal structure of a process by enumerating all possible DAGs for the relevant set of variables in Figure~\ref{fig:motiv}. The goal is to find a DAG that agrees with the data (for example, through their log-likelihood) {\em and} with the expert analyst's prior knowledge.
When we have two variables, the causal discovery problem reduces to finding out whether the variables correlate and if we find correlation, then fixing the direction of the causal arrow (or acknowledging that we cannot fix the direction). There are three possible causal models: one in which the variables are independent (with no edge) and two with a directed edge (shown in Figure~\ref{fig:motiv:twovars}). In this case, there is a statistically significant correlation, and by our modelling assumptions (here ordinary least squares linear regression) the causal model with {\sc Rg} causing {\sc NEE} is more probable. Notice that the direction of the causal arrow depends on modelling assumptions and it is often the case that based on data alone we cannot choose one model over another. As there are only two variables, the problem is computationally quite straightforward and, with our model, the statistical question is simply whether the correlation between the variables is non-zero. 

Figure~\ref{fig:motiv:threevars} shows what happens if we have three variables. Now there are in fact 25 possible causal models of which several have a high score (score being here related to the log-likelihood of the model, see Section~\ref{sec:methods}), shown above each graph.
We would expect a CSD algorithm to output one of the high-scored models conditional on some data.

When there are more than three variables, the number of possible models blows up; for 4 variables there are already 543 possible models. It therefore becomes challenging and impractical to evaluate the prior probabilities for all models in the model space, or even to go through them manually. Figure~\ref{fig:motiv:fourvars} shows the causal models found by 4 representative causal discovery algorithms, which---as often is the case---all differ from each other.

This means that a better approach would be to navigate through the space of possible models in an efficient manner. The outputs of the various causal structure discovery algorithms shown in Figure~\ref{fig:motiv:fourvars} then act as starting points for such a navigation. 

The contributions of this paper are as follows:
\begin{itemize}
\item We review the related work.
\item We show how the outputs of existing causal discovery algorithms can be used to take the expert's prior knowledge into account via interaction.
\item We show with artificially generated examples that our method is able to find a local maximum of the approximate posterior with the expert's knowledge as a prior.
\item We show with real world examples that we can improve upon results of state-of-the-art causal discovery algorithms and come by with more refined solutions that make sense for the experts, making the causal discovery algorithms more usable in practice.
\item We show with real world examples how cross-validation can be used to detect problems of overfitting and concept drift in causal analysis.
\end{itemize}

This paper is structured as follows. We review the related work in Section~\ref{sec:related}, formulate our method formally in Section~\ref{sec:methods}, demonstrate experimentally that our methods is able to capture the expert's insights and improve the results of causal discovery algorithms in Section~\ref{sec:experiments}, and conclude with discussion in Section~\ref{sec:discussion}.

\section{Related Work}\label{sec:related}

In this section, we address relevant work including those on causal structure discovery (CSD) algorithms, causal discovery in the Earth system sciences, and recent advances in interactive causal discovery.  

CSD refers to the problem of identifying causal relationships in observational data by analysing its statistical properties. There are many dedicated CSD algorithms which can broadly be categorised into constraint-based and score-based methods. Constraint-based algorithms discover DAGs based on how well they satisfy conditional independence constraints between the measured variables. Constraint-based algorithms include the PC algorithm~\citep{spirtes1991algorithm} and its variants: CPC~\citep{colombo2014order}, MPC~\citep{ramsey2006adjacency}. Fast Causal Inference (FCI) works similarly but allows for inference even in the presence of latent confounders by dropping an assumption known as "causal sufficiency"~\citep{spirtes2000causation}. Designed for detecting lagged causal relations from time series data, PCMCI also belongs to constraint-based algorithms~\citep{runge2020pcmci, runge2019detecting}.

Score-based algorithms use a scoring metric to score candidate DAGs and choose the highest scoring DAG. For example, BIC or log-likelihood of the model can be used to score the models. The search for causal models may be performed in the space of Markov equivalence classes, sets of DAGs with indistinguishable conditional dependency relationships, as in GES~\citep{spirtes2000causation} or in the space of DAGs as in FHC~\citep{gamez2007fast}. 

Recently, functional causal models (FCMs) or structural equation models (SEMs) have been used to represent non-Gaussian noise as well as non-linear relationships between variables of interest. The linear non-Gaussian acyclic model, or LiNGAM~\citep{shimizu2006linear}, the post-nonlinear (PNL) causal model~\citep{zhang2010pnl}, and the non-linear additive noise model~\citep{hoyer2008nonlinearanm} are examples of algorithms that use such an approach. For a detailed and comprehensive review of causal discovery algorithms, see~\citep{spirtes2000causation}

Given the increasing availability of large scale measurement data, CSD methods have recently been applied in the field of Earth system sciences as well---see~\citet{runge2019nature} for a review on this topic. In the Earth system sciences, Granger causality~\citep{granger1969causality} has been popular in practical research~\citep{kaufmann1997climate, kodra2011granger, smirnov2009granger}, possibly due to its clear temporality-based definition and its simple applicability. Granger causality has, however, been criticised as serving to find forecasting rather than causal relations~\citep{hamilton1994timeseries}. Studies in Earth system sciences have also used the PC algorithm~\citep{deng2014pc, ebert2012pc, ebert2015pcfci, samarasinghe2019pc} and a comparison study tested its variants, LiNGAM, and variants of GES~\citep{liu2020causalcomparison}. Non-linear additive noise modelling~\citep{hoyer2008nonlinearanm} has been applied on multiple bivariate problems in the field of geoscience and remote sensing~\citep{perez2019anm}. The FCI algorithm has been used to detect possible latent variables~\citep{samarasinghe2018fci}, although its computational cost has been proposed as one reason for the scarcity of practical applications~\citep{ebert2015pcfci}. Specifically designed for causal analysis of time series data, PCMCI has been applied to detect causal connections from data sets with a temporal dimension~\citep{krich2020, nowack2020pcmci}.

The methods listed above are all algorithmic and depend on the statistical assumptions made by the CSD methods being true in the application context. Taking into account an expert's judgment about applicable assumptions and their priors about the true generating processes has the potential to improve the performance of a CSD method. Eliciting causal beliefs from domain experts in the DAG setting is time-consuming and challenging.
Yet, there have been non-interactive approaches to using prior knowledge together with CSD algorithms~\citep{meek1995ges, odonnell2006camml, scheines1998tetrad, wallace1996camml}.
Such an approach has been applied, for example, in medicine~\citep{flores2011incorporating} and atmospheric science~\citep{kennett2001camml}. 

There have been some efforts related to interactive causal structure discovery.
Outcome-Explorer~\citep{hoque2021outcomeexplorer} allows for a causal DAG to be specified via a combination of CSD methods and user interaction: the user is able to interactively edit the presence or direction of the edges of a discovered DAG. The approach is focused on allowing users to interactively understand the causal relations of the model by changing values of the nodes and applying interventions.

Visual Causality Analyst (VCA) provides similar support for finding causal models with a fixed PC-style algorithm to provide the initial model~\citep{wang2015visual}. Multiple causal models can be built for separate subsets of the input data with the Causal Structure Investigator~\citep{wang2017visual}, a continuation on VCA. SeqCausal is a similar approach designed for causal structure discovery for event sequence data from multiple sources~\citep{jin2020visual}.

None of the interactive CSD methods above show the user multiple possible initial models simultaneously to choose from or what effect different edits would have on the model fit. They also do not include validation of the found model to evaluate the model fit and to detect problems such as overfitting and concept drift.
Cross-validation has been applied in causal discovery in the algorithm Out-of-Sample Causal Tuning (OCT) to perform algorithm selection and hyperparameter optimisation~\citep{biza2020tuning}.
We propose using validation for model scoring when the expert user navigates in the space of causal models.

Gathering information from domain experts is not trivial~\citep{garthwaite2005statistical} and the elicited information is prone to a multitude of biases~\citep{tversky1974uncertainty}. However, rather than eliciting prior probability distributions, we ask the expert to incorporate their beliefs regarding the conditional independences between the variables in the model. This has been stated to be a simpler and more straightforward task than probability elicitation~\citep{garthwaite2005statistical}. Explicitly stating the assumptions brought into the model by the expert further alleviates the issues of uncertainty. If all of the included assumptions are known, they can be scrutinised after a model has been found. Listing the assumptions also enables replication of the obtained results. Furthermore, assumptions made during the model discovery may provide ideas about which experiments should be performed in order to reduce uncertainty over the model.

\section{Methods}\label{sec:methods}

In this section, we introduce a theoretical formalisation of interactive causal structure discovery with an expert user and describe our practical implementation.
We propose that interactive causal structure discovery should comprise of obtaining a selection of possible initial models, navigating in the space of causal models, and using cross-validation to detect overfitting and concept drift.
We do not aim to give a definitive answer to how these should be formalised or implemented but present one way of manifesting the theory.
For example, we assume the user to be a rational Bayesian agent with a constant prior, but of course more complex formulations would be possible.

\subsection{Formulation}

Given a data set $X$, the task is to find a model that fits $X$ and agrees with the user's prior knowledge. We formulate the problem using a Bayesian approach.
We assume that we are given a likelihood $p(X\mid \theta)$ that can be computed and that the user has a prior $p_U(\theta)$ which is not known. $p_U(\theta)$ encodes the expert's knowledge on present and absent causal relations and directions. The objective then is to find the user's maximum a posteriori (MAP) solution $\theta_U = \arg \max_{\theta} p(X \mid \theta) p_U(\theta)$. 

Since the user's prior $p_U(\theta)$ is not known, finding the user's MAP solution is not trivial. We have chosen to model the user's solution using a greedy search with user interactions. The user starts at an \emph{initial state} $\theta_1$ and is allowed to make local moves in the parameter space into its \emph{neighboring states} $N(\theta)$ (defined below). We assume the user moves greedily to states with higher probability in the user's posterior, eventually resulting in a local MAP solution.
Thus, at iteration $t$, the next state is given by
\begin{align*}
    \theta_{t + 1} = \arg \max_{\theta \in N(\theta_t)} p(X \mid \theta) p_U(\theta)
\end{align*}
Once there are no more moves that increase the posterior given the user's prior, the user stops the exploration. With this process, $\theta_U$ or at least a local optimum of the user's posterior is found. 

We next describe how the above formulation applies to graphical models specified as DAGs. We parameterise a DAG as $(\theta, \beta)$, where $\theta$ represents parameters about the DAG structure as edge probabilities and $\beta$ represents parameters about modelling assumptions, such as functional forms of parent-child relationships, regression coefficients, and noise distributions. Writing their joint distribution as $p(\theta, \beta) = p(\beta | \theta) p(\theta)$ allows us to specify separately a prior over the structure and a prior over the model parameters given the structure. 

For the purposes of this paper, we define the neighbourhood $N(\theta)$ of a DAG $\theta$ to be the DAGs that are one edit distance away. Neighboring states are therefore reached by making an \emph{edit} to the current state by either adding, removing, or reversing an edge in the DAG. The initial state $\theta_1$ is obtained from a CSD algorithm. This provides us with a simple cognitive model that we can use later to model the interaction of the user with the causal discovery system.

It is helpful to consider a special case where the CSD algorithm is Bayesian in nature (which it of course doesn't have to be!). Assume that the Bayesian CSD algorithm uses a known prior distribution $p_C(\theta)$ (the ``computer prior'') to find the best model $\theta_C$ and the best model (output by the CSD algorithm) would be given by the MAP solution $\theta_C = \arg \max_{\theta} p(X \mid \theta) p_C(\theta)$ where $\beta$ has been integrated over, $p(X \mid \theta) = \int p(X \mid \theta, \beta) p_C(\theta) d\beta$, assuming the prior $p_C(\theta)$ for $\theta$.
The MAP solution found by the user may be different than the one found by the computer if the computer prior and user prior differ.
Conversely, if we were able to elicit the user prior and use it in the CSD algorithm then the algorithm would directly output the user's MAP solution.

\subsection{Implementation}\label{subsec:implementation}

The interactive CSD process is implemented using a graphical user interface. 
The process begins with the expert selecting an initial model from the DAGs output by multiple CSD algorithms. 
The DAGs are displayed together with their respective model scores and, for the currently selected model, the change in score is shown for every possible edit: addition, removal, or reversal of an edge.
Using the presented information together with their prior knowledge, the expert may then choose to edit the current model in order to navigate to a neighboring model. 
The edits are performed by clicking on an upper-triangular adjacency matrix corresponding to the current model.
All performed edits and model scores are stored and shown to the expert throughout the process. 

As a model score, we use an averaged adjusted coefficient of determination, or $\overline{R}_{a}^2$, over all of the $I$ variables in the model, which is essentially a scaled log-likelihood of the model under the assumptions of Gaussianity and linearity.
The advantage of using $R^2$ rather than the model's raw estimated log-likelihood to measure goodness-of-fit stems from its easy interpretation as the proportion of variance explained, which is a common measure in the Earth system sciences.
The model score $\overline{R}_a^2$ is computed as follows: each variable is linearly regressed on its parents, an adjusted coefficient of determination ($R_a^2$) is computed for that regression model and, finally, the mean of the computed values is returned as the full causal model's score.
Since at least one variable in a DAG has no parents and thus has $R_a^2 = 0$, the range of the simple average over the adjusted coefficients of determination is $[0, (I - 1) / I]$.
By multiplying the mean value by $I / (I - 1)$, we obtain $\overline{R}_a^2$ with range $[0, 1]$ for training data
\begin{equation*}
 \overline{R}_a^2 = \frac{1}{I - 1} \sum_{i=1}^I R_{i,a}^2
 = \frac{1}{I - 1} \sum_{i=1}^I \left( 1 - (1-R_i^2) \frac{N-1}{N-\mid {\rm Pa}(X_i) \mid-1} \right),
\end{equation*}
where $N$ is the sample size, Pa($X_i$) the set of parents of variable $X_i$, and $R_i^2$ is the coefficient of determination when regressing $X_i$ on its parents.
When there are exactly two variables, the $\overline{R}_a^2$ matches the traditional definition of the adjusted $R^2$ score and it is proportional to the log-likelihood of the model under the assumptions of linearity and normally distributed noise. 

We compute $\overline{R}_a^2$ on a training-validation split to estimate and communicate possible overfit and concept drift to the user.
With the validation scores for the current model and its neighbouring models, the user can make navigational decisions that in part affect which causal model is found in the end.
For independent data, we form a training and validation set by randomly sampling two equal-sized sets.
For time series data, we use blocked cross-validation~\citep{bergmeir2012cv}: the data are split into temporally contiguous blocks, each of which is used for validation in one iteration while the rest are used for training. 
The final validation score is computed as an average over the validation scores for the blocks.
The validation score for each block is computed by training the regression models on the training set and then computing the $\overline{R}_a^2$ for the validation set. 
Note that $\overline{R}_a^2$ can be negative in the validation set, since the model predictions used to compute individual $R_{i,a}^2$ are from models trained on a separate data set. 
In such a case, the interpretation is that the mean value of the validation data set provides a better prediction than the trained model.

\section{Experiments}\label{sec:experiments}

In this section, we perform experiments on synthetic data using a simulated user and present use cases on real-world data with and without expert knowledge. We first describe the algorithms and data sets used in the experiments, and then present the experimental setups and the results for both simulated user experiments and real world use cases. The experiments were performed using R~\citep{rmanual}, version 3.6.3, and the source code is available online.\footnote{\url{https://github.com/edahelsinki/ICSD}}

\subsection{Initial Models for Navigation}
In our approach, CSD algorithms are used to provide initial models for the expert user to begin their analysis.
The algorithms' outputs can also act as ``global'' navigation points, instead of local navigation with single edits.
The algorithms included in the experiments comprise PC-Stable with two significance levels 0.1 and 0.01, GES, and ICA-based LiNGAM.
The main reasons for selecting these algorithms were to have a diverse group of algorithms based on differing assumptions for which ready implementations exist.
The approach can easily be extended to include other algorithms for which reason the set included at this stage is somewhat irrelevant.
All of the selected algorithms assume causal sufficiency and linear causal relations.
Other algorithms we considered, FHC~\citep{gamez2007fast} and FCI~\citep{spirtes2000causation}, were not included, as the run time of FHC was too long for the experiments given our resources and the results from FCI would not be comparable: FCI outputs a partial ancestral graph instead of a DAG or a Markov equivalence class output by the other chosen algorithms.
We used the implementations available in the R package \verb|pcalg|~\citep{kalisch2012pcalg} for the CSD algorithms.

\subsection{Data Sets}

The synthetic data set is created by generating a random directed acyclic graph and then sampling the graph with random edge weights for data sets of varying sizes.
Each graph is generated with a sparsity of 0.3: each pair of variables has an edge between them with a probability of 0.3.
Acyclicity is ensured by orienting all edges in the order the variables are defined, away from the first variable.
The noise for each variable follows a zero-mean distribution which is randomly chosen from two options: either uniform distribution (-0.01, 0.01) or Gaussian with a standard deviation of 0.01.
The reason for including both types of noise distributions is to create data sets which almost follow assumptions made by the algorithms while still breaking some of them.
We tested creating data with different amounts of noise but that had no significant impact on the results.
All of the algorithms we use in the experiments assume linearity but, additionally, PC-Stable and GES assume Gaussianity of noise and LiNGAM assumes non-Gaussianity.

The real-world data set consists of measurements collected at the SMEAR II (System for Measuring Forest Ecosystem-Atmosphere Relationships II) station at Hyyti\"al\"a, Finland, which is located in a Scots Pine forest, regenerated by sowing in 1963 after clear-cut. The data used here were collected from 2013 to 2015~\citep{hyyfluxnet}, when the dominant trees were about 17 m tall.
The data are part of the FLUXNET2015 data set~\citep{pastorello2020fluxnet}.
The measurement data were averaged at half hour intervals.
Variables included in the analysis are shortwave downward radiation (Rg), air temperature (T), vapour pressure deficit (VPD), sensible heat flux (H), latent heat flux (LE), and net ecosystem exchange (NEE).
Rg is the solar radiation in the wavelength range 0.3-4.8 $\mu$m. Air temperature is measured at 8 m height above ground, and vapour pressure deficit is calculated based on this temperature and the measured relative humidity. Ecosystem scale sensible and latent heat fluxes, which are related to surface-atmosphere dry air heat transfer and energy flux related to evapotranspiration, respectively, as well as NEE were measured above the forest canopy using the eddy-covariance (EC) technique~\citep{rebmann2018ec}. EC fluxes were calculated using EddyUH software~\citep{mammarella2016ec} according to standard methodologies~\citep{sabbatini2018ec}.
%\todo{Tuomo and Ivan: Short description of the variables used - Rg, T, VPD, H, LE, NEE}
We keep observations where the potential shortwave downward radiation is at least 80\% of the daily maximum to mitigate effects of diurnal variation on the data distribution~\citep{krich2020}.
Additionally, measurements with gap filled values for NEE, H, or LE are filtered out to avoid introducing artificial causal relations.
After filtering, the data set contains 817 data points for April 2013-2015, 854 data points for May 2013-2015, and 215 data points for August 2015.
We perform the analysis on three month combinations: April, April \& May, and April \& August.

\subsection{Simulated User Experiments}

To analyse an interactive CSD process, we simulate a user parameterised by $k \in [1/3, 1/2]$ that represents their level of knowledge.
Each pair of variables has three possible states in terms of causal dependence: not connected or connected with an arrow in either direction.
The user prior for an edge thus consists of a discrete distribution with three exclusive events.
As we know the true state of each edge, $k$ determines the prior probability of the true state of an edge and the two remaining states have prior probabilities of $(1-k)/2$ each.
If $k=1$, the user knows all edges in the causal graph that generated the synthetic data with probability one, in which case the posterior is dominated fully by the prior.
If $k=1/3$, the user has no prior information about the causal structure and the posterior is dominated by the model likelihood.
Generally, $k$ can take values in $[0,1]$ but we do not take into account wrong information, $k < 1/3$, and values above $1/2$ do not produce interesting results as such high certainty leads to near-constant results.

We also test the effect of partial knowledge where the user has information regarding two thirds of the pairs of variables but no knowledge of the remaining pairs.
This corresponds to using two values of $k$, one for the known parts of the graph and another, $1/3$, for the unknown parts.

We compare models using the structural Hamming distance (SHD)~\citep{dejongh2009comparison}.
The SHD between two graphs represents the number of edits required to transform one graph into the other.
Each edit comprises adding, deleting, or removing an edge.

For each set of parameters ($k$, sample size, number of variables, amount of knowledge), a hundred random graphs are generated to find meaningful distributions for the results.

\begin{figure}[t]
 \centering
 \begin{subfigure}[b]{\textwidth}
    \centering
    \includegraphics[width=\columnwidth]{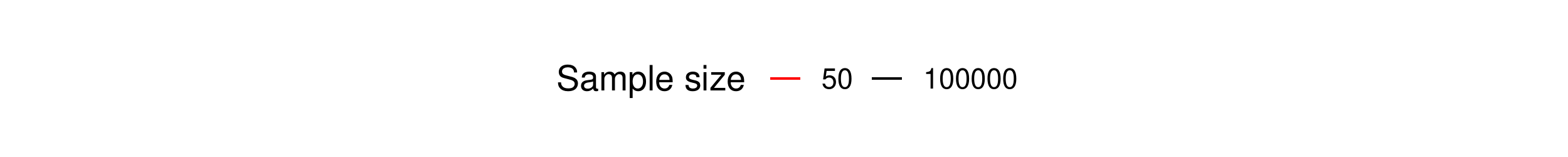}
 \end{subfigure}
 \begin{subfigure}[b]{.49\textwidth}
    \includegraphics[width=\columnwidth]{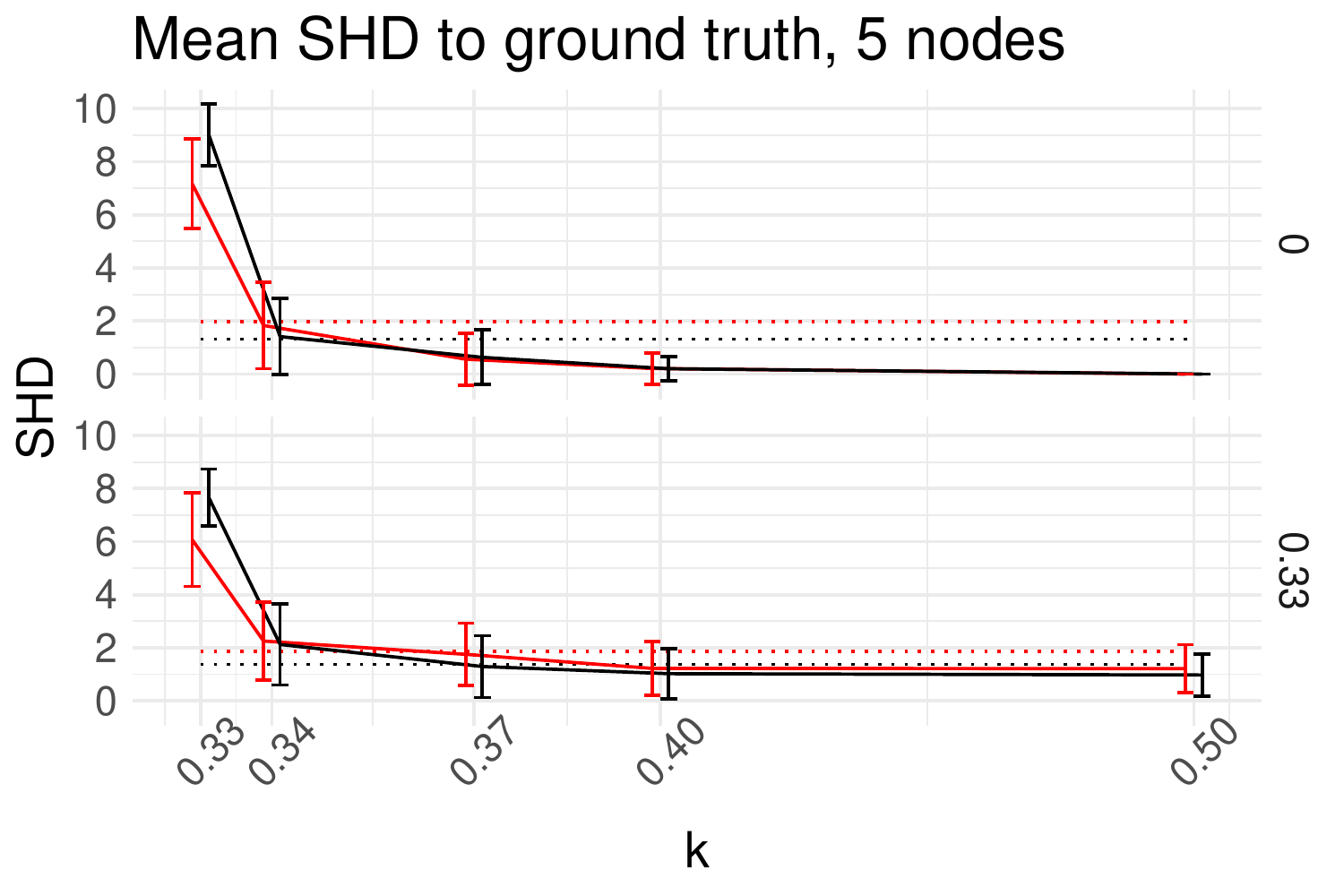}
    \caption{Results for graphs with 5 nodes.}
    \label{fig:shd:single:smallgraph}
 \end{subfigure}
 \begin{subfigure}[b]{.49\textwidth}
    \includegraphics[width=\columnwidth]{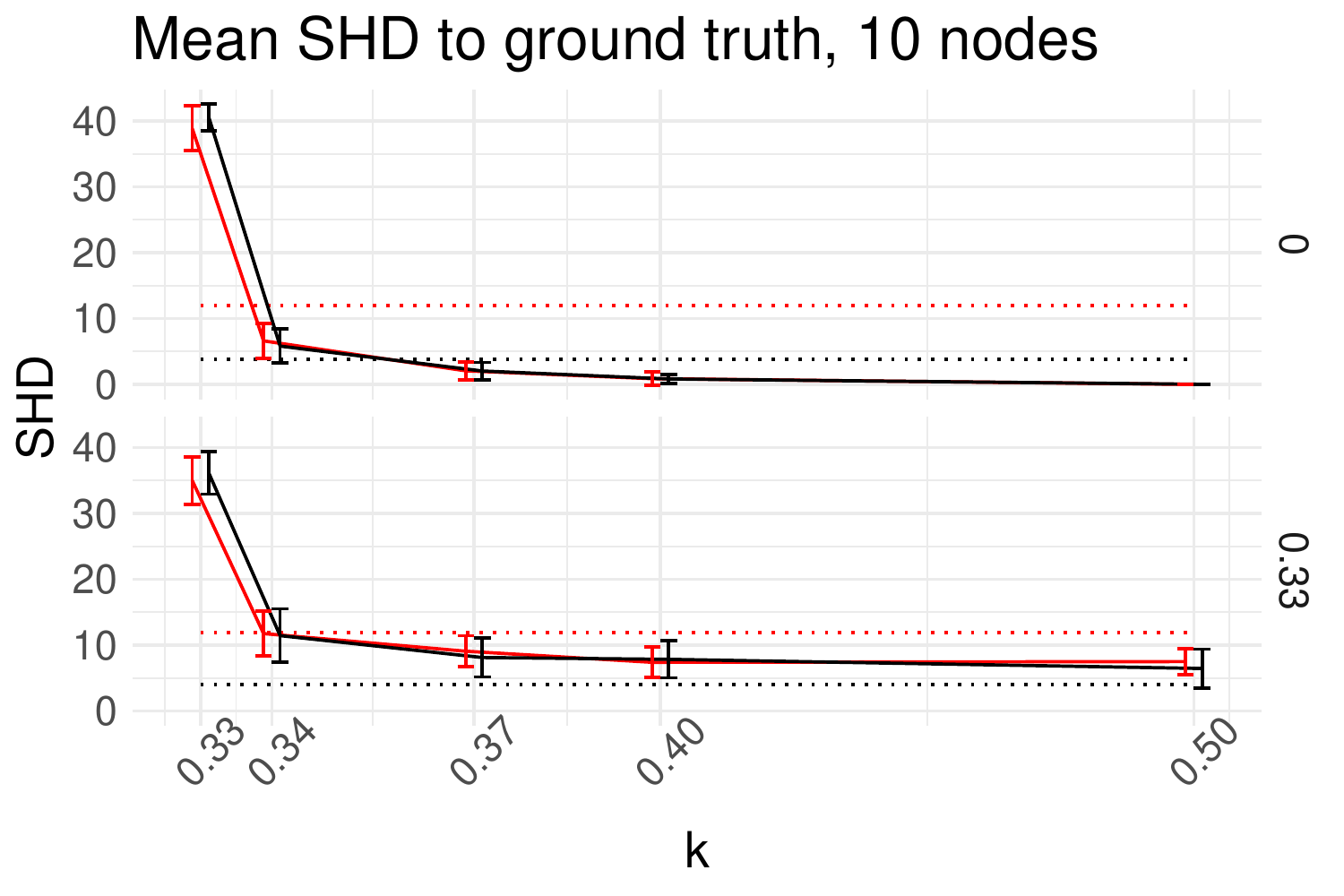}
    \caption{Results for graphs with 10 nodes.}
    \label{fig:shd:single:largegraph}
 \end{subfigure}
 \caption{Experiment 1. Mean structural Hamming distances to the ground truth model. Error bars represent mean $\pm$ one standard deviation, dotted lines the mean SHD of the initial model to ground truth. Above results with full knowledge, below one third of knowledge missing. Incorporating expert knowledge leads to models closer to ground truth, especially with small sample sizes and high level of knowledge.}
 \label{fig:shd:single}
\end{figure}

\subsubsection{Experiment 1: Does Incorporating Expert Knowledge into the Search Result in Better Models?}

In Experiment 1, we examine how expert knowledge results in better models by simulating a user navigating in the model space. 
The highest scoring output from the default set of CSD algorithms is chosen as the initial model.
Then, the model is edited one step at a time, greedily selecting the neighbouring model with the highest user posterior.
For each model, the posterior is computed using the simulated user's prior, parameterised by $k$, combined with the model's approximate log-likelihood.
When the current model has the highest posterior over all its neighbours, the navigation ends.
The final model is compared with the ground truth model using SHD.

Figure~\ref{fig:shd:single} shows the results of Experiment 1. 
We see that with knowledge on all pairs of variables (upper figures), higher user knowledge leads to improvements over the initial model, denoted by a dotted line.
In contrast, greedily optimising the model score, which corresponds to using a flat user prior with $k = 1/3$, generally leads to worse models than the initial model in terms of SHD.
This is because under a non-uniform prior, the true model does not necessarily have the highest $\overline{R}_a^2$ score which is proportional to the log-likelihood.
For example, with 10 nodes and small sample size, the average score of the true model is 0.54 while the average for a model found with interaction is 0.66.
For 5 nodes and large samples, the corresponding values are 0.37 for the true model and 0.44 for the result of navigation.
Final models obtained by using a flat prior differ from the initial models because the initial model may not have the highest score which results in the simulated user navigating greedily to models with higher scores.
The results underline the need to rely on both the data-based score and expert knowledge to find good models.

Even when the user has no knowledge of the causal connections between a third of the variable pairs (bottom figures), user interaction improves the initial model when there is little data or few variables.
When the expert's knowledge has no missing information (upper figures), the resulting models are closer to the true model than the initial model already for $k \geq 0.34$ with small sample size and for $k \geq 0.37$ with large samples. This suggests that even minimal user knowledge leads to better models.

As expected, increasing the amount of data improves the performance of the CSD algorithms leading to better initial models, seen as the black horizontal being always below the red horizontal. 
However, a high level of knowledge still improves these initial models converging them towards the true model which, in these experiments, corresponds to the posterior model of the simulated user.
Increasing the number of variables (Figure~\ref{fig:shd:single:largegraph}) increases the model space which negatively affects the initial model given by the CSD algorithms, leading to higher SHD values.
Again, including user knowledge still improves the final result in all cases except in the case of large data and missing knowledge (bottom Figure~\ref{fig:shd:single:largegraph}). 

\begin{figure}[t]
 \centering
 \includegraphics[width=0.85\columnwidth]{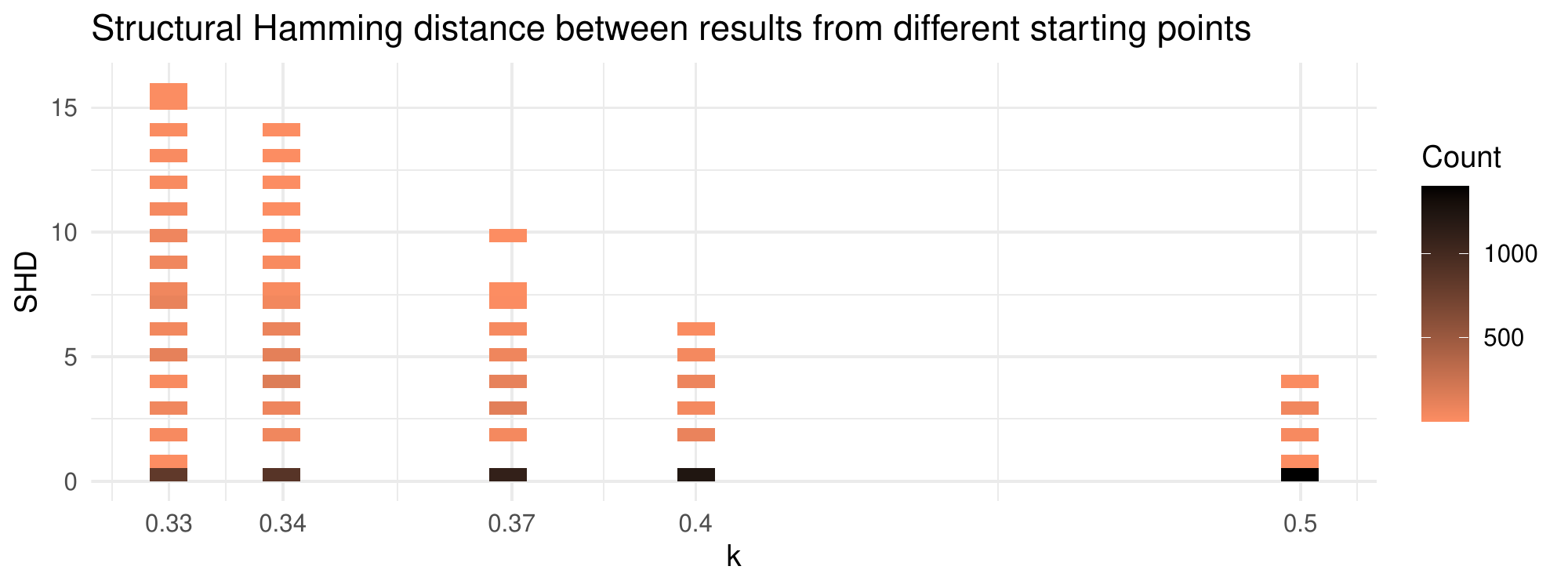}
 \caption{Experiment 2. Pairwise structural Hamming distances when running analysis on the same data starting from different initial models. Variance in the distances shows the final model is affected by choice of initial model.}
 \label{fig:multidist}
\end{figure}

\subsubsection{Experiment 2: Is It Useful to Have Algorithms Provide Initial Models?}

In Experiment 2, six graphs are used as initial models for the navigation: an empty graph, the true graph, and the highest scoring model for each of the four default algorithms.
Navigation is performed as in Experiment 1 but the resulting models are compared using SHD with each other instead of the ground truth.
Comparing models with each other allows us to determine whether the initial model affects which model is found and, therefore, whether it is useful to have a selection of different initial models.

Figure~\ref{fig:multidist} shows the pairwise SHD between final models when using different initial models for the same data. 
The resulting final models are mostly similar with most pairwise SHD values at zero but, with lower levels of knowledge, there is more variance in the results.
This is expected: if the expert has strong knowledge of the underlying data generating process, the initial model bears little importance as the strong prior affects the posterior more than the likelihood.
With lower $k$, there is more uncertainty in which local optimum, of which there may be multiple, is found in the navigation.
Navigation in the space of causal models may be initiated from any graph, for example always using an empty DAG as the initial model. 
The results, however, suggest that the choice of initial model affects the result and starting from an empty graph may not produce optimal results in every case.

\subsection{Use Cases With Real World Data}

We present here three examples of interactive causal structure discovery with real-world data. In each case, the initial model is the highest scoring model output by the default set of four algorithms, PC-Stable with significance levels 0.01 and 0.1, GES, and LiNGAM.
We compute training and validation scores as described in Section~\ref{subsec:implementation}.

\begin{figure}[t]
    \centering
    \includegraphics[width=\textwidth]{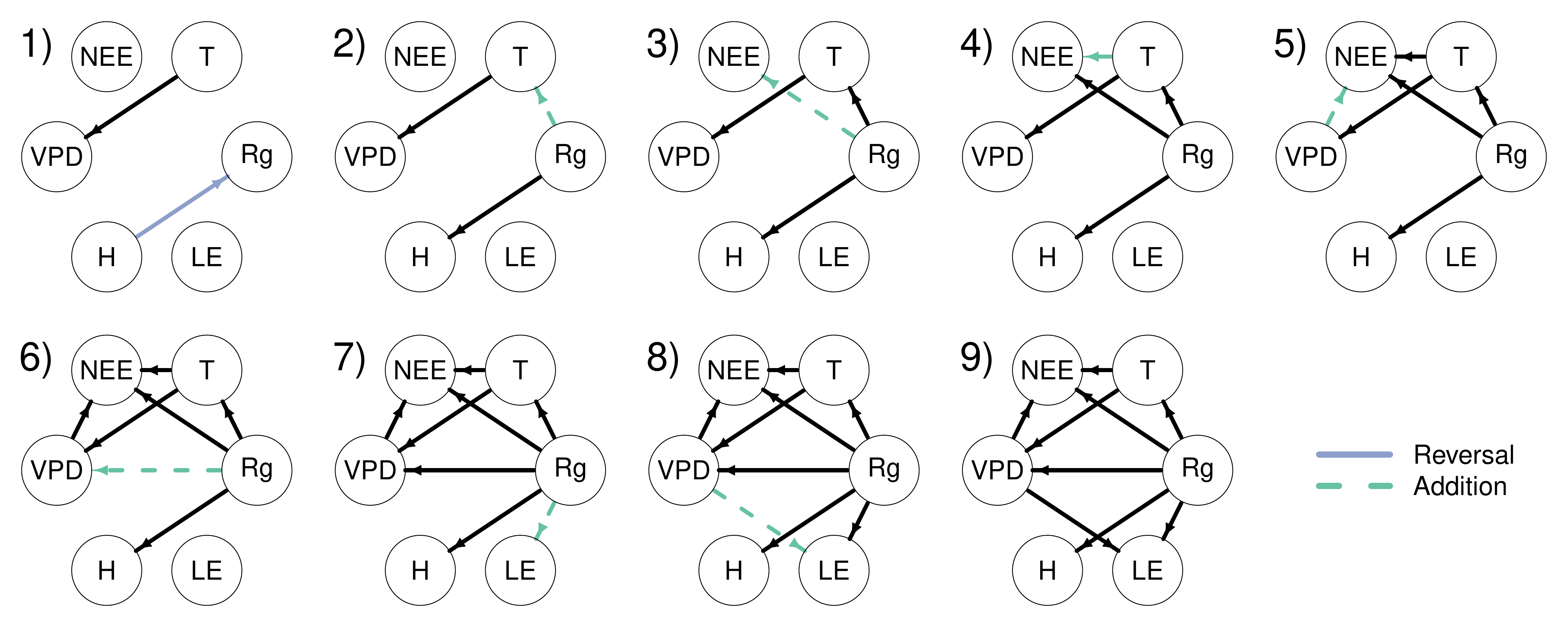}
    \caption{An example of a user navigation using data for April 2014.}
    \label{fig:expert:navigation}
\end{figure}

Figure~\ref{fig:expert:navigation} displays an example of how an expert user may edit a graph. The user starts modifying the initial model 1 shown in Figure~\ref{fig:expert:navigation} by adding a connection from downward shortwave radiation to temperature and sensible heat flux (model 2). This is justified, since the solar radiation affects the ambient air temperature as well as heats up the ground. In the next steps (models 3 and 4), connections from Rg and T to NEE are added. In our boreal forest site, Rg is the driving factor for photosynthetic activity of the plants and T is controlling the soil and plant respiration~\citep{markkanen2001eddycovariance}. VPD is connected to NEE in model 5, since VPD may affect the plants' CO$_2$ exchange due to the opening and closing of the stomata according to the amount of water vapour in the air. The heating of ground and other moist surfaces by solar radiation can lead to increase in water evaporation, therefore a connection from Rg to LE is added in model 6. Changes in the amount of water vapour present cause changes in the evaporation rate of water from surfaces, therefore in model 7 a connection from VPD to LE is added. The final model 8 is the combination of the models 1-7. The trajectory of model scores through the navigation are shown in Figure~\ref{fig:expert:april}.

\subsubsection{Use Case 1: Detection of Overfitting}

\begin{figure}[t]
    \centering
    \begin{subfigure}[b]{\textwidth}
        \centering
        \includegraphics[width=\textwidth]{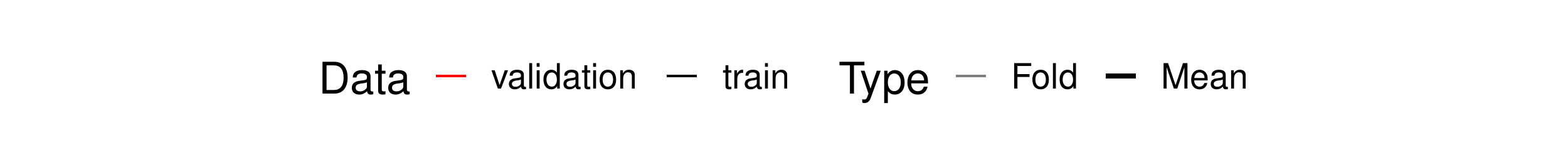}
    \end{subfigure}
    \begin{subfigure}[b]{0.48\textwidth}
        \includegraphics[width=\textwidth]{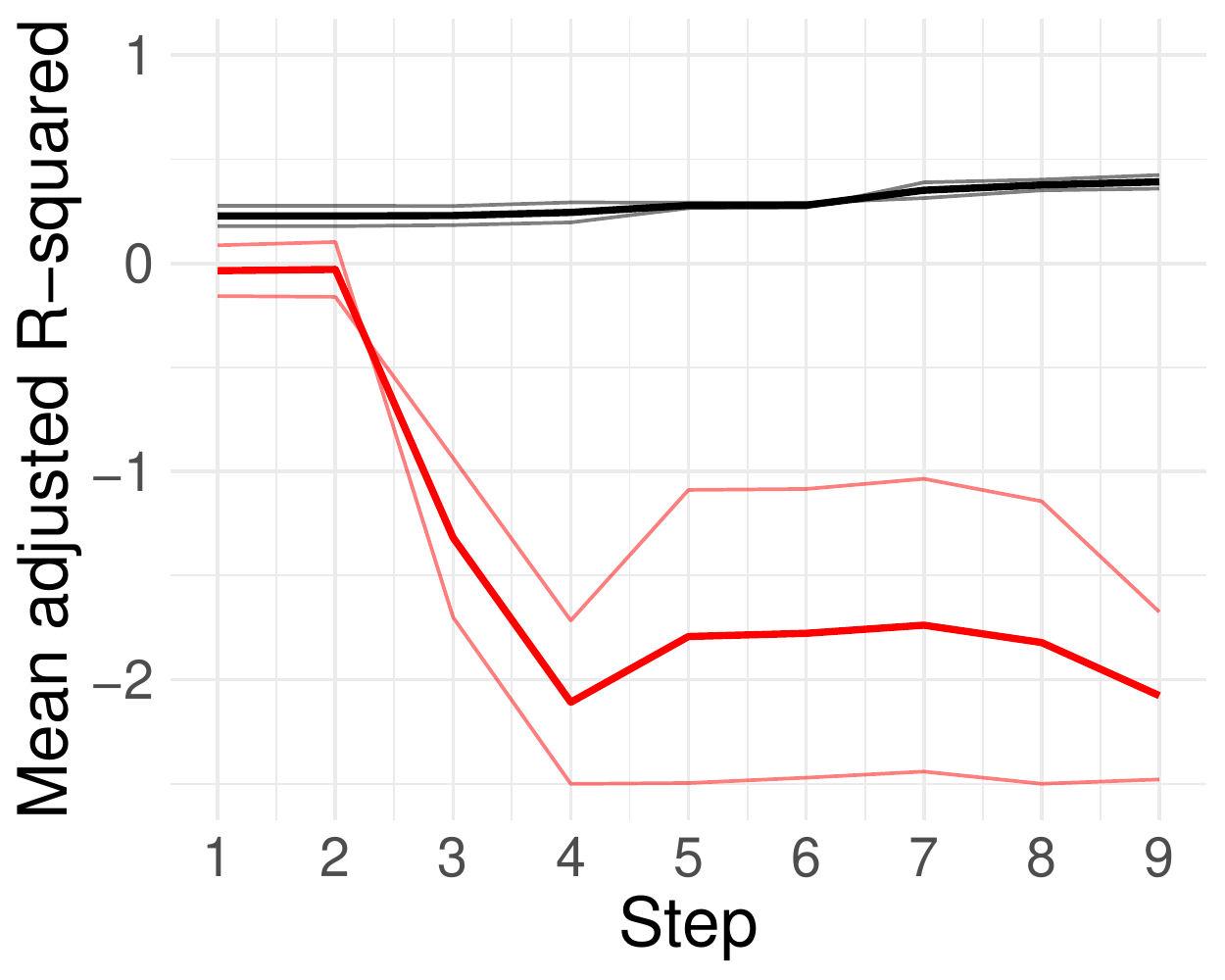}
        \caption{}
        \label{fig:expert:april}
    \end{subfigure}
    \hfill
    \begin{subfigure}[b]{0.48\textwidth}
        \includegraphics[width=\textwidth]{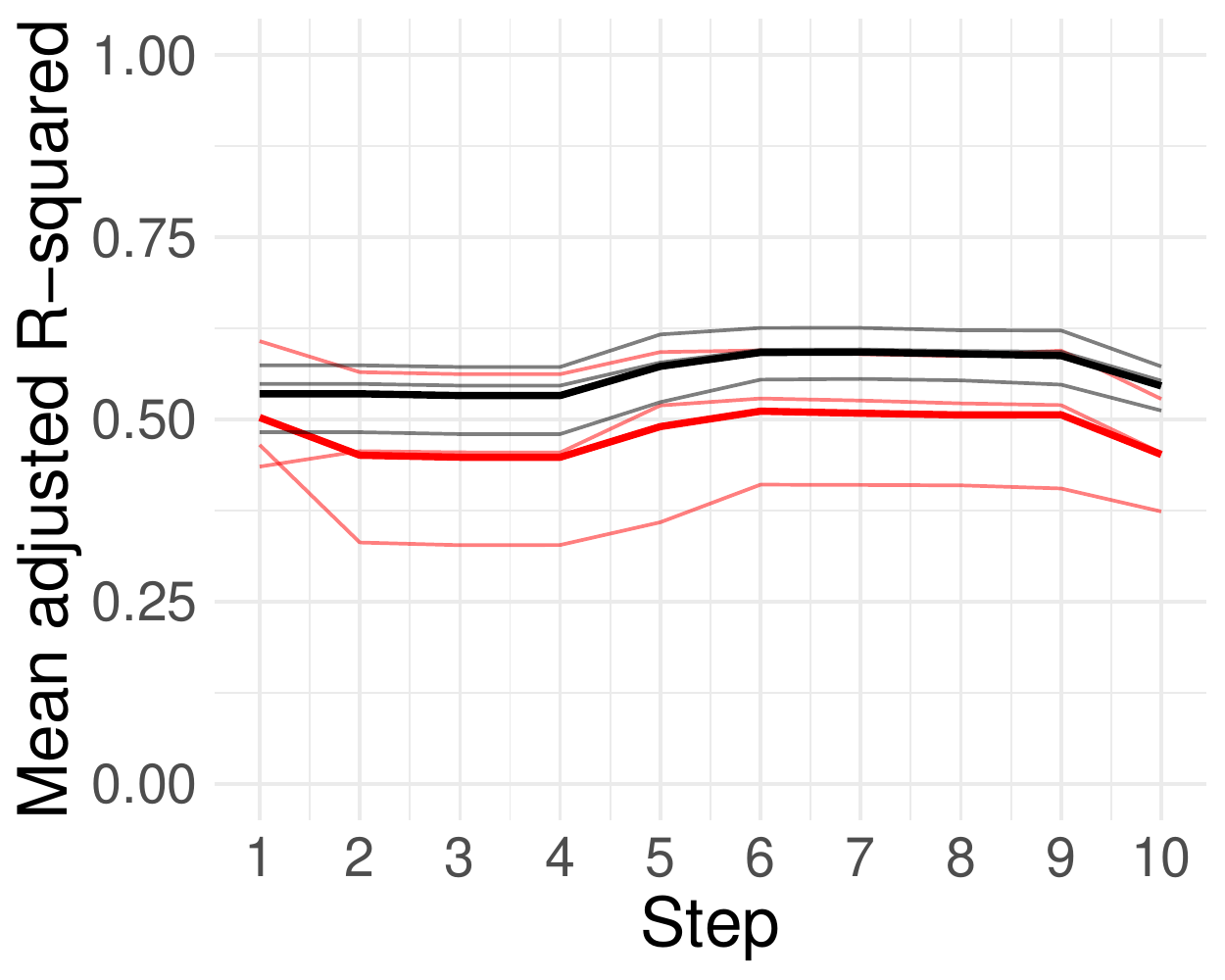}
        \caption{}
        \label{fig:expert:aprilmay}
    \end{subfigure}
    \caption{Use case 1. Showing the user the validation and training scores allows them to (a) detect and (b) fix overfitting problems. Validation and training $\overline{R}_a^2$ with data from (a) April 2014; (b) April and May 2013--2015. 
    Note the different scaling of the y-axes.}
    \label{fig:expert:overfit}
\end{figure}

Overfitting is a common problem in modelling although, to the best of our knowledge, it has not been addressed in previous work in the context of interactive causal structure discovery.
In this use case, we demonstrate how the user may detect overfitting by inspecting the training and validation scores, and differences between them using 2-fold blocked CV on data measured in April 2014.
Already the initial model, the best model output by the default algorithms according to the $\overline{R}_a^2$, has a negative validation score which is shown in Figure~\ref{fig:expert:april}.
As discussed in Section~\ref{subsec:implementation}, a negative validation score indicates the mean of the validation data produces better predictions than the trained model.
After the model is edited through interactions, the training and validation scores diverge radically.
Negative validation scores throughout the navigation may indicate overfitting or concept drift, and further investigation is required to determine the cause of the problem.
Because the data contain samples from one month only, a likely issue is overfitting: the model specialises on the training data leading to inability to predict the validation data well.
Once the user adds more data to cover both April and May in 2013 through 2015, the validation score stays clearly positive for all three cross-validation folds and the training and validation score averages follow the same pattern through the interactions, as shown in Figure~\ref{fig:expert:aprilmay}. 

Cross-validation is a well-known technique for controlling overfit, although it has not been applied in interactive CSD, to the best of our knowledge.
This use case shows how it may be used in interaction with the user. 
Without checking models for overfitting, we risk obtaining a model that does not generalise or does not reflect the true phenomena in the data.
Showing the user validation and training scores enables them to decide whether the model has overfit the data or not.

\subsubsection{Use Case 2: Detection of Concept Drift}

\begin{figure}[t]
    \centering
    \begin{subfigure}[b]{\textwidth}
        \centering
        \includegraphics[width=\textwidth]{img/UseCases/R2_legend.pdf}
    \end{subfigure}
    \begin{subfigure}[b]{0.48\textwidth}
        \centering
        \includegraphics[width=\textwidth]{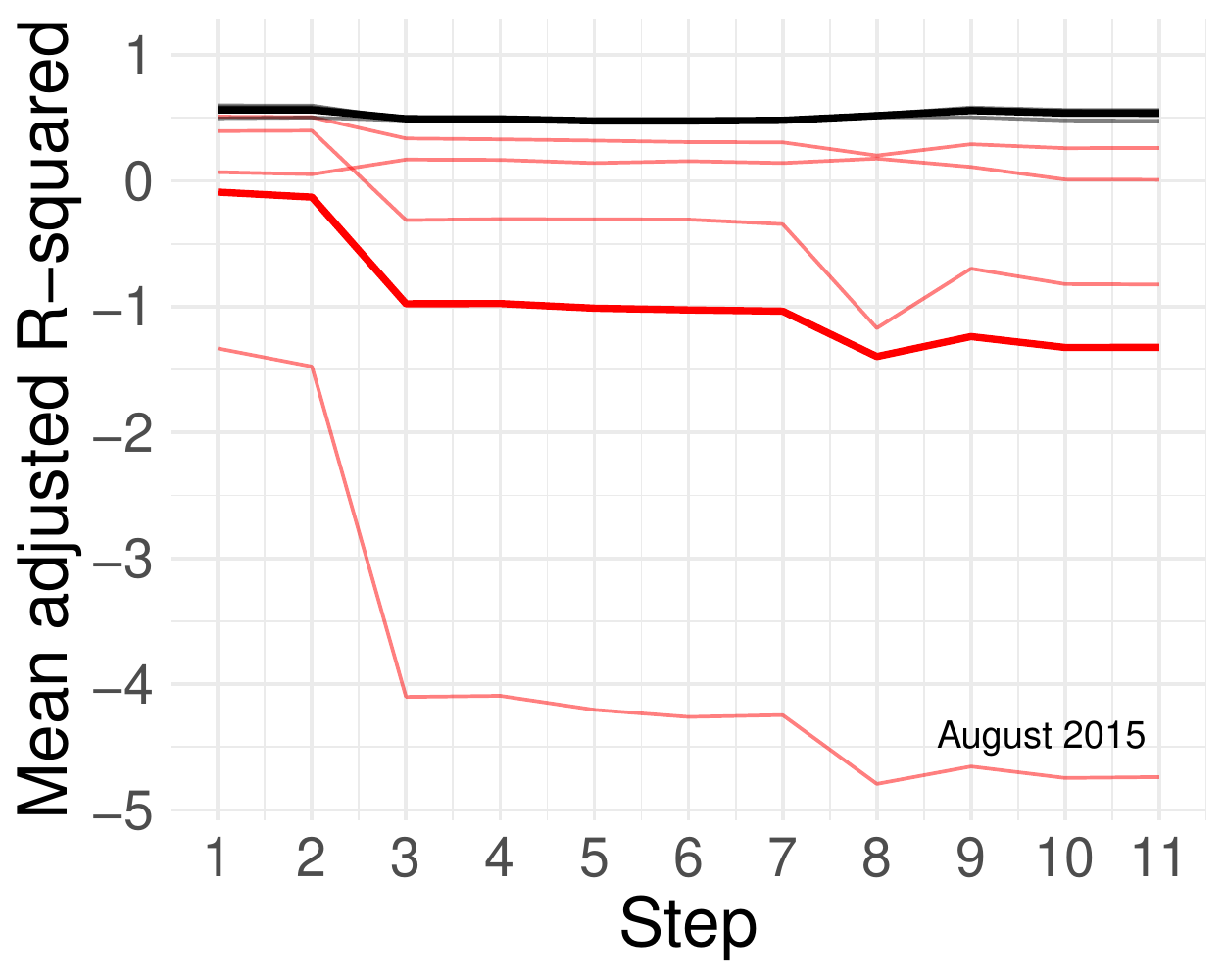}
        \caption{}
        \label{fig:expert:aprilaugust}
    \end{subfigure}
    \hfill
    \begin{subfigure}[b]{0.48\textwidth}
        \centering
        \includegraphics[width=\textwidth]{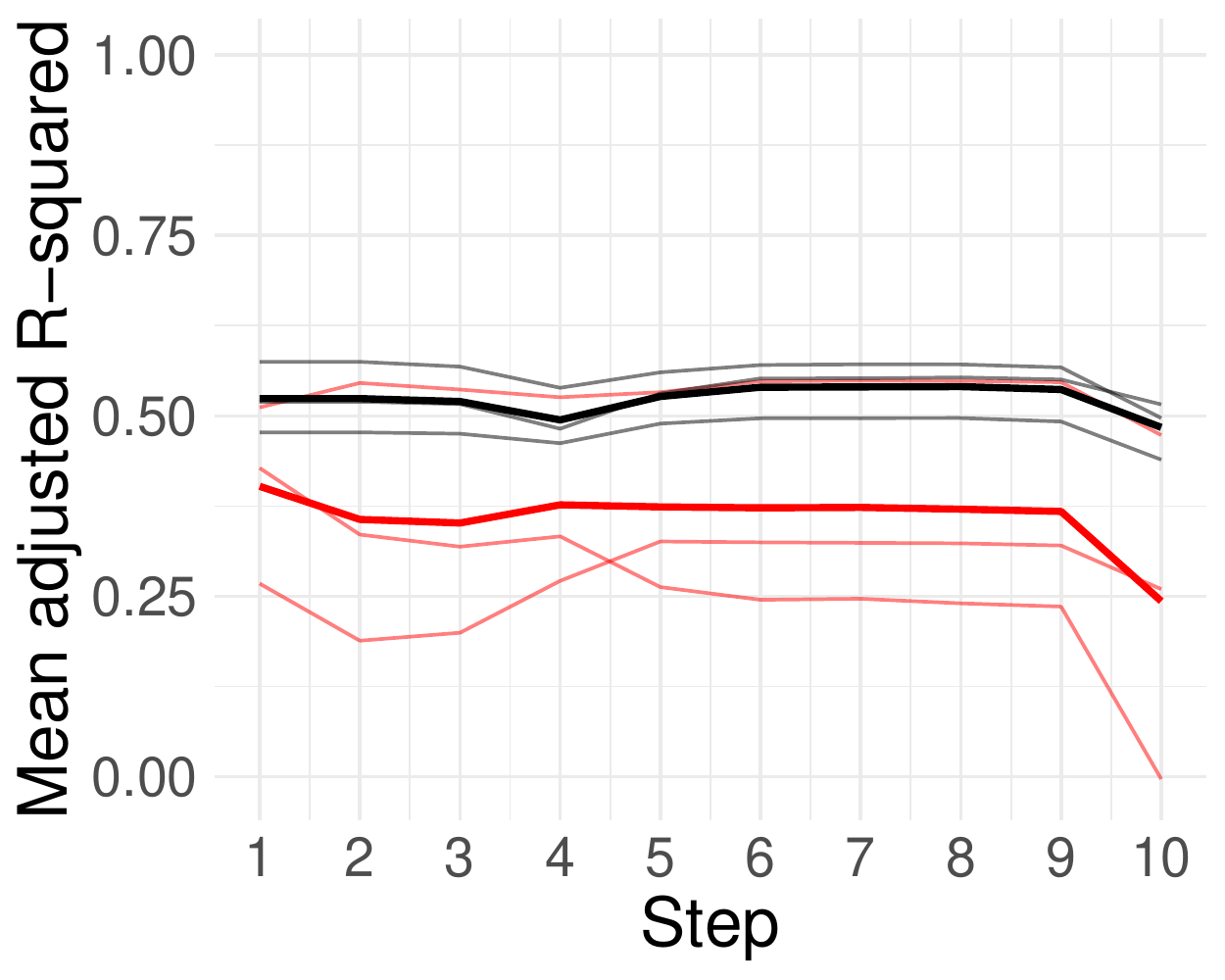}
        \caption{}
        \label{fig:expert:aprillong}
    \end{subfigure}
    \caption{Use case 2. The user can (a) detect concept drift and (b) remove data points with a different generating distribution to improve model fit. Validation and training $\overline{R}_a^2$ with data from (a) April 2013--2015 and August 2015, (b) April 2013--2015.  Note the different scaling of the y-axes.}
    \label{fig:expert:conceptdrift}
\end{figure}

Another problem that can arise in causal modelling with real-world data is concept drift.
To demonstrate how the user can detect concept drift with blocked cross-validation, we analyse a data set containing samples from April in 2013--2015 and from August 2015.
Similarly to the previous use case, the validation score falls below zero after two edits and for one of the folds, the score is negative already for the initial model.
Furthermore, the validation and training scores start diverging after the first edit.
We notice that although the validation score is negative for more than one of the folds, the score for the fold containing the August data is clearly inferior to the rest, which suggests potential concept drift.
Removing the problematic August data improves the scores significantly, leading to similar, non-negative trajectories for the training and validation scores, as shown in Figure~\ref{fig:expert:aprillong}. August 2015 was very dry and warm in our measurement site, leading to high VPD and low soil water content. These non-optimal conditions for the photosynthetic activity of the plants are likely the reason for the different causal connections between the studied variables in August 2015 compared to data collected in April 2013--2015.

This use case demonstrates how the user can detect concept drift that may occur in real-world systems.
Undetected concept drift may result in a model that fits none of the similarly distributed subsets of the data well.
Problematic subsets of the data can be identified by the user with information on the validation and training scores for each of the cross-validation folds consisting of contiguous data blocks.

\subsubsection{Use Case 3: Effect of Initial Model}

\begin{figure}[t]
    \centering
    \begin{subfigure}[b]{0.49\textwidth}
        \centering
        \includegraphics[width=0.6\textwidth]{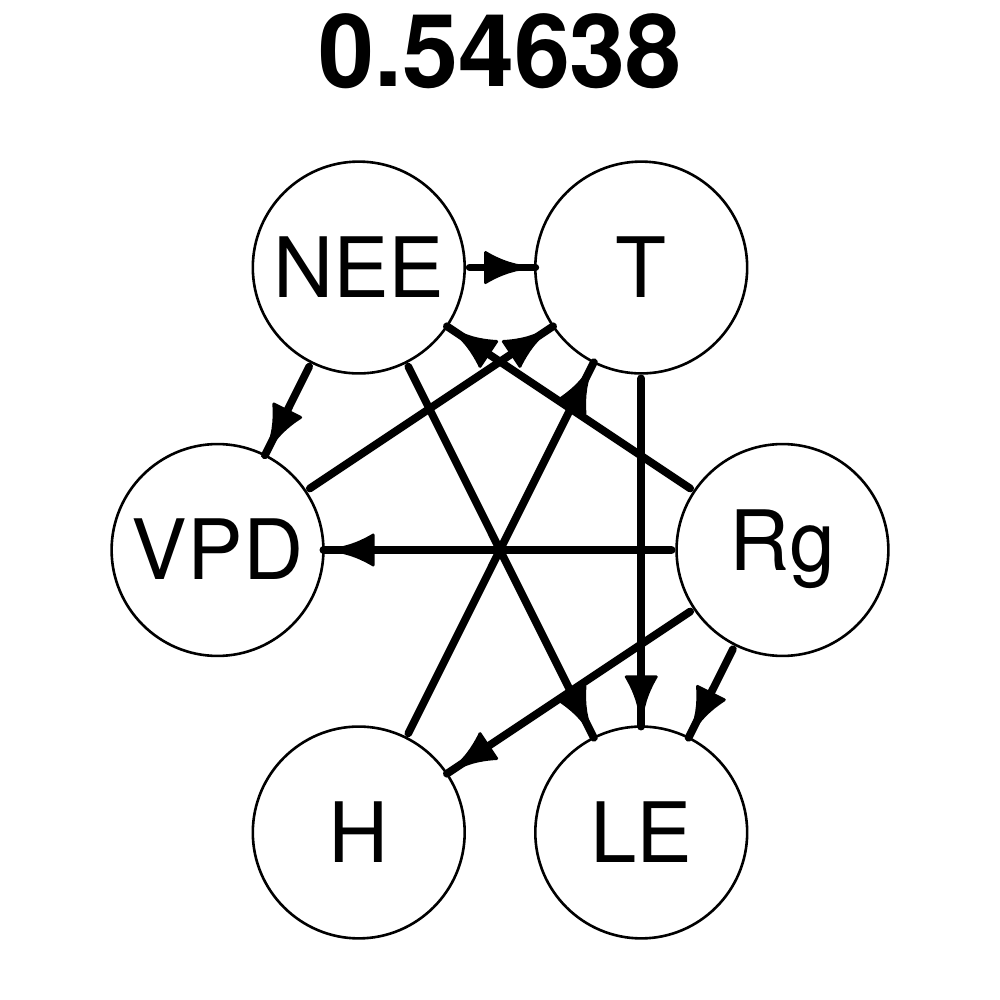}
        \caption{}
        \label{fig:non-expert:alg}
    \end{subfigure}
    \hfill
    \begin{subfigure}[b]{0.49\textwidth}
        \centering
        \includegraphics[width=0.6\textwidth]{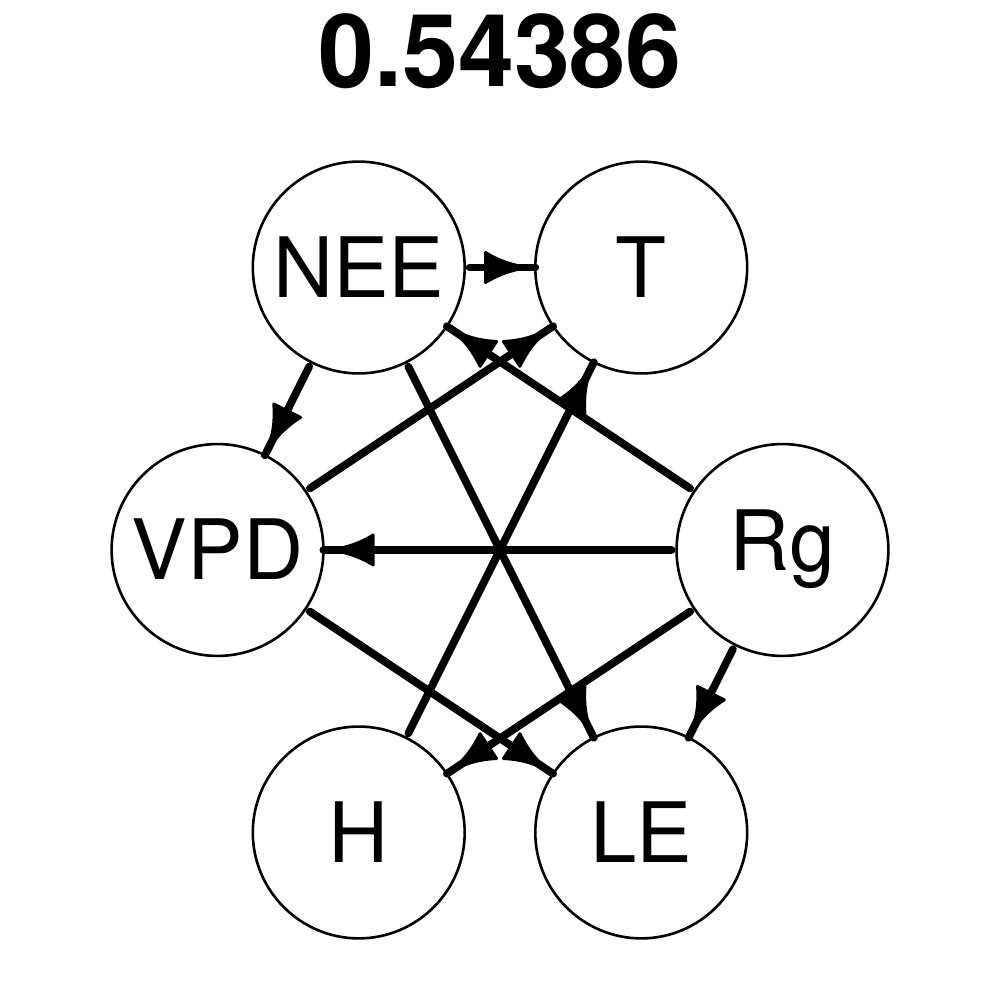}
        \caption{}
        \label{fig:non-expert:empty}
    \end{subfigure}
    \caption{Use case 3. Result of navigation without any prior knowledge of the model when starting from (a) the algorithm output with highest score and (b) an empty graph. While similar, the final models are not equal with different initial models.}
    \label{fig:non-expert:models}
\end{figure}

Depending on the choice of initial model, different models that fit the data may be found.
When the expert has knowledge of causal relationships between all pairs of variables, the initial model has little impact on the final model as the strong prior dominates the posterior.
However, when the user has little or no knowledge of the data generating process, results are more sensitive to variations in the initial model due to the greedy approach to finding a local optimum of the posterior.
To demonstrate this, we assume a uniform user prior and begin navigation both from the highest-scoring output obtained by the default set of CSD algorithms and from an empty graph.
In each step, we greedily navigate to the neighbour with highest score and stop once the score cannot be improved by a single edit.
The final models obtained with no knowledge and different initial models are displayed in Figure~\ref{fig:non-expert:models}.
A slightly different local optimum of the approximate posterior is reached depending on the initial model.
Although the two models are quite similar with a structural Hamming distance of just two, the example highlights the possibility of finding a different model that fits the data equally well or better when changing the initial model.
For the curious reader, we also note here that the structural Hamming distance between the experts' model 8 in Figure~\ref{fig:expert:navigation} and the final models in Figure~\ref{fig:non-expert:models} are seven and five, respectively.

\section{Discussion}\label{sec:discussion}

In this paper, we have presented a principled procedure for studying interactive causal structure discovery (ICSD). 
We view ICSD as a user navigating in the space of possible DAGs, or a combinatorial optimisation problem in which the optimiser is the expert user. 
Unlike the field of causal structure discovery (CSD), which has a long history and where many algorithms have been proposed, the field of ICSD is still nascent and there are many open challenges. 

The motivation for the paper stems from applying CSD algorithms in the Earth system sciences. If an expert user runs multiple CSD algorithms on the same data, they obtain multiple causal models. This can be confusing and it is often not clear which assumptions underlie the output models or if the models could be modified to take the user's domain knowledge into account.

We claim that the raw outputs from CSD algorithms need to be edited by an expert user to obtain valid causal models. We proposed a formalisation of the interactive procedure in which the expert edits a given DAG, and we used a simulated user to study the interactive CSD process. The results suggest that even small levels of prior knowledge are useful in improving the outputs from CSD algorithms with user interaction. We also demonstrated in the use cases how overfitting and concept drift can occur and be detected in ICSD. Prior work in ICSD has not considered overfitting and concept drift, and has instead relied on the user to regularise the process. We proposed cross-validation as a means for detecting and communicating overfitting and concept drift to the user in ICSD.

Our current formulation takes a greedy approach which leads to local optima. This was partly shown in our experiments, where the choice of the initial state affects the final model. Better final models may be found by providing several initial states. The initial states here were graphs from multiple CSD algorithms, but they may be sampled in other ways, such as Markov Chain Monte Carlo methods~\citep{friedman2000beingbayesian, viinikka2020mcmc} and stability selection~\citep{meinshausen2010stability, stekhoven2012stability}.

In addition to allowing navigation through interactions, expert knowledge could be incorporated already in the initial causal discovery with MCMC and stability selection as well as other CSD algorithms used to obtain the initial models. How the results are affected by incorporating expert knowledge in the initial CSD algorithms, through interactions, or both, remains a topic for future research together with comparisons among different methods of obtaining the initial models and different cross-validation methods. Inspecting separately the SHD between models caused by existence of causal relations and by orientations might provide further insight into the results.

Our work, while providing a working solution, is meant to highlight issues encountered and point out avenues for future research in order to make CSD algorithms truly usable in Earth sciences and similar fields, where the data are interpreted by experts with a deep understanding of the processes involved.
We recognise there are multiple alternatives to our approach for incorporating expert knowledge in CSD, such as using Bayesian priors over variable orderings~\citep{friedman2000beingbayesian}, and detailed comparison with related work remains a topic for future research.
We finally discuss the open research questions that we find particularly important and interesting.

{\bf User model.} We modelled the user as a rational Bayesian agent, with simplifications such as implicitly assuming that the user's prior knowledge stays constant. Obviously, this model can be only a crude approximation of the reality. We did not take cognitive biases or other limitations into account. {\em What would be a better model and would it have impact on the actual implementations? How could we generalise it to a collaborative setting where there are several experts? How could we test the validity of the user model for this particular task?}

{\bf Causal model representation.} We represented the causal model by DAGs, but we did not directly address effect sizes, which may be important in practice: we can have statistically, yet not practically significant correlation. A crucial part in our approach is to show the user how well the model fits the data; we used an $R^2$ measure, or re-scaled log-likelihood, for this task. In order for the user to make informed navigation choices, they should have a good understanding of the likelihood and do the ``mental computation'' needed to choose a step that maximises the posterior probability, expressed as the sum of log-likelihood and the user's prior. Also, in this work we used a simple linear model to compute the log-likelihoods. {\em What would be a good way to show effect sizes? How should we describe to the user the fit of the model to the data? What modelling assumptions, aside from linear, could we use to compute the effect sizes?}

{\bf Starting points for exploration.} Now we used outputs of several CSD algorithms for the exploration. This may not be optimal. Intuitively, we would like to have a set of starting points that would cover all local optima: a global MAP solution could be found by starting from at least one of the proposed starting points. {\em How could we find a representative set of starting points for the exploration?}

{\bf Integration of interaction into workflow.} The current causal modelling tools offer only limited support for interactive model building~\citep{gelman2020}. In order for ICSD to be practical, the software tools should implement the interactive workflow.

{\bf Evaluation of interactive modelling methods.} Introducing a user into the modelling workflow complicates the evaluation of such a system. We examined a simple user model but more complex evaluation methods are possible. {\em What would be a better benchmark for an interactive CSD method? What should the objectives and evaluation metrics?}

% Acknowledgements should go at the end, before appendices and references

\acks{We thank Helsinki Institute for Information Technology, Future Makers Funding Program, and Finnish Center for Artificial Intelligence for support.}

% Manual newpage inserted to improve layout of sample file - not
% needed in general before appendices/bibliography.

% \newpage

% \appendix
% \section*{Appendix A.}
% \label{app:theorem}

% Note: in this sample, the section number is hard-coded in. Following
% proper LaTeX conventions, it should properly be coded as a reference:

%In this appendix we prove the following theorem from
%Section~\ref{sec:textree-generalization}:

\vskip 0.2in
\bibliography{references}

\end{document}